\documentclass[aps,pra,twocolumn,showpacs]{revtex4}
\usepackage{amsmath}
\usepackage{epsfig}
\usepackage{bm}
\usepackage{epsfig}

\def\be{\begin{equation}}
\def\ee{\end{equation}}
\def\bea{\begin{eqnarray}}
\def\eea{\end{eqnarray}}
\def\la{\langle}
\def\ra{\rangle}

\begin{document}

\title{Fluctuation statistics of mesoscopic Bose-Einstein condensate: reconciling
the master equation with the partition function to revisit the
Uhlenbeck-Einstein dilemma}
\author{Andrew N. Jordan}
\thanks{All authors contributed equally to this work, and are listed alphabetically.}
\affiliation{Department of Physics and Astronomy, University of Rochester, Rochester, New York 14627, USA }
\affiliation{Institute for Quantum Studies and Department of
Physics, Texas A\&M University, TX 77843-4242}
\author{C.~H. Raymond Ooi}
\thanks{All authors contributed equally to this work, and are listed alphabetically.}
\affiliation{Institute for Quantum Studies and Department of
Physics, Texas A\&M University, TX 77843-4242}
\author{Anatoly A. Svidzinsky}
\thanks{All authors contributed equally to this work, and are listed alphabetically.}
\affiliation{Institute for Quantum Studies and Department of
Physics, Texas A\&M University, TX 77843-4242}
\date{August 10, 2006}

\begin{abstract}
The atom fluctuations statistics of an ideal, mesoscopic, Bose-Einstein
condensate is investigated from several different perspectives. By
generalizing the grand canonical analysis (applied to the canonical
ensemble problem), we obtain a self-consistent equation for the mean condensate
particle number that coincides with the microscopic result calculated from the
laser master equation approach. For the case of a harmonic trap, we obtain
an analytic expression for the condensate particle number that is very
accurate at all temperatures, when compared with numerical canonical
ensemble results. Applying a similar generalized grand canonical treatment
to the variance, we obtain an accurate result only below the critical
temperature. Analytic results are found for all higher moments of the fluctuation 
distribution by employing the stochastic path integral formalism, with excellent 
accuracy. We further discuss a hybrid treatment, which weds the 
master equation/stochastic path integral analysis with the results obtained 
based on canonical ensemble quasiparticle formalism [V. V. Kocharovsky \textit{et al.}, 
Phys. Rev. A \textbf{61}, 053606 (2000)], producing
essentially perfect agreement with numerical simulation at all temperatures.
\end{abstract}


\maketitle

\section{Introduction}

The problem of fluctuation statistics in mesoscopic Bose-Einstein condensate
(BEC) is a rich one - even, somewhat surprisingly, for the noninteracting
Bose gas, as well as for the interacting Bose gas \cite{Koch06}. Take, for
example, a trapped gas of $N$ bosons, and lower the temperature below the
critical temperature $T_c$ to form a condensate. Count the number of atoms
that are in the condensate. Do this many times at the same temperature and
total boson number $N$ to get statistics. What is the distribution of the
condensed atoms, and how does it change when moving the temperature through $%
T_c$? The noninteracting, mesoscopic, Bose gas is a problem whose
fluctuation characteristics are still not completely understood. Contrary to
standard lore, we stress that the fluctuations are not Gaussian, even in the
thermodynamic limit. The discussion goes clear back to the
Uhlenbeck-Einstein dilemma \cite{Uhlenbeck,EinsteinBEC}: Uhlenbeck's
criticism that Einstein's expression for the average boson number $\langle
n_0 \rangle$ had an abrupt cusp at $T_c$. The manner in which this cusp is
smoothed by fluctuations is of great interest for mesoscopic systems, which
is one focus of the present paper. Even more subtle is the question of the
higher order moments of the condensate fluctuations, especially in the
vicinity of $T_c$.

Despite the conceptual simplicity of the above question, the fluctuation
statistics are not known analytically, because while the canonical ensemble
partition function can be formally written, it can be accurately calculated
only numerically. We note that the work by Holthaus and Kalinowski obtains
accurate approximations for the first few moments, employing a refined
saddle point approximation of the canonical partition function \cite{sp}.
However, the equation for the saddle point still has to be solved
numerically.

\begin{figure*}[t]
\center\epsfxsize=18cm\epsffile{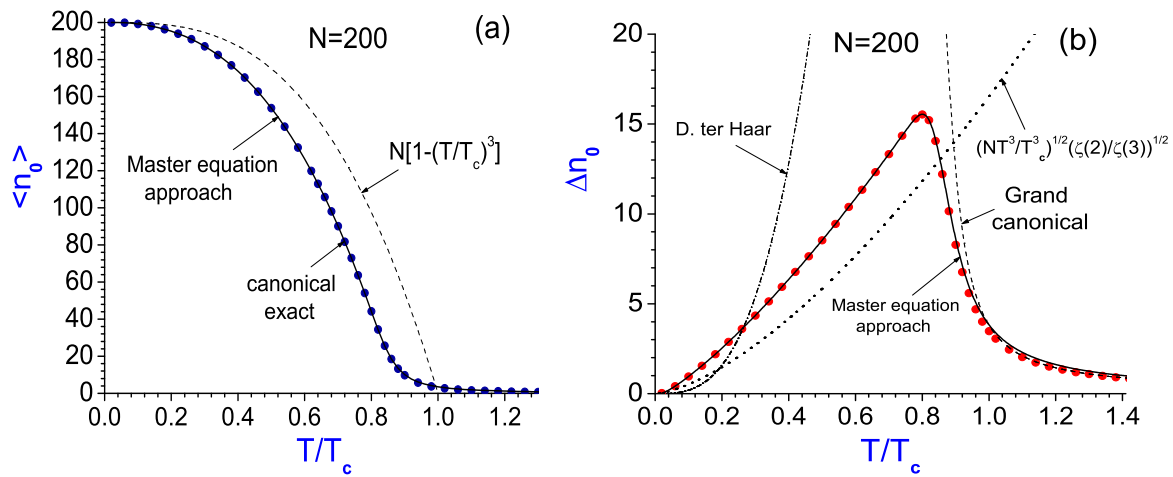} \caption{(Color online) (Left). The
average boson number in the condensate is plotted versus
temperature.   The Master equation approach (solid line) is
compared with numerical simulation of canonical results (dots), as
well as the thermodynamic limit (dashed line). (Right). The
variance of the distribution is plotted versus temperature.
The grand canonical result (dashed line) deviates near and below the
critical temperature, leading to the ``grand canonical
catastrophe".  The result of ter Haar \cite{th} (dash dot) and
Politzer \cite{pol} (small dots) (derived in the thermodynamic
limit) are also shown. The results are shown for $N=200$ particles
in a harmonic trap.} \label{cond}
\end{figure*}

The time-dependent master equation is an excellent alternative approach to
attack this question, developed in the pioneering papers of Scully \cite
{CNBI} (hereafter referred to as CNB1) and Kocharovsky, Scully, Zhu and
Zubairy \cite{CNBII} (hereafter referred to as CNB2). This approach is
capable of investigating dynamical phenomena, as well as being a viable
option when we include interactions, or when the system is out of
equilibrium. The master equation approach, within the canonical framework,
has successfully calculated moments of the fluctuation distribution to good
accuracy. Other advantages of this method are that it provides a simple
physical picture of the condensation process with a gas of cold atoms via a
detailed balance of the elementary processes of heating and cooling, and
that the theory shows a vivid analogy between atom BEC and the photon laser
by demonstrating that the BEC and laser master equations are identical. This
analogy can explain the atom-laser linewidth \cite{linewidth}, and might
also be able to explain the transient dynamics of the phase transition to
BEC. In the master equation approach, the cooling of the BEC is accomplished
by the non-condensate interaction with a thermal environment. In typical BEC
experiments, the cooling is accomplished with evaporative techniques.
However, in the experiments by Reppy and co-workers \cite{reppy}, the BEC
was cooled by a thermal reservoir associated with the Vycor glass host --
exactly the situation we presently consider. The master equation takes a
given bath model for the calculation, so one natural question is if a
different bath model alters the physics of the BEC statistics. 
While different models will certainly alter the dynamics of the condensation, we
expect the statistics to be robust, 
just as the analogous partition function
is.

How does all of this relate to traditional statistical physics, in
particular the microcanonical, canonical, and grand canonical
formalism?
Recall that Einstein predicted BEC by analyzing the grand canonical formula
\cite{Einstein}. Both grand canonical ensemble and canonical ensemble
approaches describe the occurrence of BEC equally well in the dramatic rise
of the mean condensate number $\langle n_{0}\rangle $ in the ground state.
However, near the critical temperature and below, the grand canonical
approach gives grossly different results for the variance and higher
moments, the so-called ``grand canonical catastrophe'' illustrated in Fig.
\ref{cond}b.

The question of fluctuations in Bose-Einstein condensation has attracted
renewed interest. Grossmann and Holthaus \cite{GHcusp,GHlambda} previously
obtained an analytical expression for the mean condensate particle number in a
harmonic trap for finite $N$ using the grand canonical approach. However,
their analytical expression does not show a smooth crossover near $T_{c}$.
Grossmann and Holthaus \cite{GHcusp} and Ketterle and Druten \cite
{KetterleDruten} found the crossover behavior only by numerical solution of
the equation for the condensate particle number. Politzer also obtains an
approximate expression for the variance in the thermodynamic limit \cite{pol}.

The purpose of the present paper is to report advances on several fronts.
One goal is to connect the master equation formalism with the partition
function analysis in thermal equilibrium. To this end, we review the master
equation formalism, and derive an approximate result for the mean condensate
particle number, that can be expressed in terms of elementary functions. This result
is then connected with standard statistical mechanics by extending the grand
canonical approach (applied to the canonical ensemble problem!) to obtain an
approximate result for the mean condensate particle number that is quite accurate
through the critical temperature. This is done by deriving a self-consistent
equation for the condensate particle number by eliminating the chemical
potential. The equation may be approximately solved, and the solution
exactly coincides with the approximate master equation result.

Next, we investigate fluctuations, a more subtle problem. The analogous
treatment for the variance via the generalized grand canonical analysis
works quite well for low temperatures, but fails for temperatures near or
greater than the critical temperature. In order to obtain a more accurate
analytic result from the master equation approach, we first reformulated the
master equation as a stochastic path integral. The stochastic path integral
formalism \cite{anj1,anj2} allows for an accurate approximation of all
cumulants of the fluctuation statistics in terms of elementary functions, by
applying the saddle-point approximation, whose large parameter is the number
of condensed atoms. It is straightforward to generalize the results, and
present the fluctuation statistics of a general boson master equation, with
arbitrary heating and cooling coefficients.

Finally, by combining the master equation/stochastic path integral
predictions with the previous results of V. V. Kocharovsky \textit{et al.}
based on canonical ensemble quasiparticle
formalism \cite{CNBIII} (hereafter referred to
as CNB3), which is very accurate at low temperature, we are able to
formulate a hybrid theory, whose results are in near-perfect agreement with
exact numerical simulation. The idea is to use the low-$T$ results of CNB3
to fix the values of three master equation parameters via the first three
moments. Then the master equation predictions for several higher moments and
cumulants are compared with the numerical results at all temperatures. They
are in near-perfect agreement.

The paper is organized as follows. In Sec. II we review the master equation
approach, and derive an approximate solution for the mean atom number in the
condensate. This solution is connected with standard partition function
methods in Sec. III by extending the grand canonical analysis, and deriving
a self-consistent equation for the mean atom number, which coincides with
the result of Sec. II. In Sec. IV, we consider a harmonic trap, and give
improved analytic results for the heating coefficient. We go on to apply the
extended grand canonical analysis to the variance of the distribution. In
Sec. V, we give the stochastic path integral treatment of the problem, which
yields approximate expression for all cumulants in terms of the master
equation parameters. In Sec. VI, we introduce a hybrid theory, combining the
predictions of the master equation/stochastic path integral, with those of
CNB3, to obtain excellent agreement with numerical simulation. We conclude
in Sec. VII. Appendices A and B contain the details of the calculations for
the grand canonical approach, and appendix C contains the details of the
stochastic path integral calculations.

\section{Master Equation Approach}

Let us review the results of the master equation approach. Consider a
trapped Bose gas, whose single-particle energy levels are $\epsilon _{k}$.
The master equation formalism begins with the elementary transitions of
atoms between the condensate and non-condensate. We are primarily interested
in the condensate statistics, and therefore focus on the probability $%
p_{n_{0}}$ of having $n_{0}$ atoms in the condensate, given that there are $%
N $ total atoms. The master equation describing the heating and cooling
processes of a Bose-Einstein condensate is given by \cite{CNBI}
%\begin{widetext}
\begin{eqnarray}
{\dot{p}_{n_{0}}} &=&-\kappa \lbrack
K_{n_{0}}(n_{0}+1)p_{n_{0}}-K_{n_{0}-1}n_{0}p_{n_{0}-1}  \notag \\
&+&H_{n_{0}}n_{0}p_{n_{0}}-H_{n_{0}+1}(n_{0}+1)p_{n_{0}+1}],
\label{generalme}
\end{eqnarray}
%\end{widetext}
where in the low temperature limit
\begin{equation}
K_{n_{0}}=N-n_{0},\qquad H_{n_{0}}=\mathcal{H},  \label{hk1}
\end{equation}
\begin{equation}
\mathcal{H}=\sum_{k>0}\frac{1}{e^{\beta \epsilon _{k}}-1},
\end{equation}
and $\kappa $ is a rate constant. The steady state solution for the
distribution is \cite{CNBI}
\begin{equation}
p_{n_{0}}=\frac{N!}{(N-n_{0})!}\frac{\mathcal{H}^{N-n_{0}}}{e^{\mathcal{H}%
}\Gamma (N+1,\mathcal{H})},
\end{equation}
where
\begin{equation}
\Gamma (N+1,x)=\int_{x}^{\infty }t^{N}e^{-t}dt
\end{equation}
is an incomplete gamma-function. Higher order moments may be found via $%
\langle n_{0}^{s}\rangle =\sum_{n_{0}}n_{0}^{s}p_{n_{0}}$. We introduce the
notation $\mu _{s}\equiv \langle (n_{0}-\bar{n}_{0})^{s}\rangle $ for the
central moments of the distribution. The equation for the probability
distribution can also be converted to an equation for the mean atom number $%
\langle n_{0}\rangle $ to find
\begin{equation}
\langle {\dot{n}_{0}}\rangle =-\kappa \lbrack (\allowbreak -N+1+\mathcal{H}%
)\langle n_{0}\rangle -N+\langle n_{0}^{2}\rangle ].
\end{equation}
Approximating $\langle n_{0}^{2}\rangle \simeq \langle n_{0}\rangle ^{2}$,
and considering the stationary case, the solution for $\langle n_{0}\rangle $
is
\begin{equation}
\langle n_{0}\rangle \simeq \frac{1}{2}\left( N-\mathcal{H}-1+\sqrt{(N-%
\mathcal{H}-1)^{2}+4N}\right) .  \label{n0meapprox}
\end{equation}
In contrast to CNB1, this approximate answer can be simply expressed in
terms of $\mathcal{H}$ and $N$, yet still catches the smooth transition in
the vicinity of $T_{c}$ (where $N-\mathcal{H}\sim \sqrt{N}$), see Fig. \ref
{meanfig} below. We will show in the next section that this same
(approximate) result can be derived within the context of the grand
canonical formalism, and later in Sec. V, how to derive similar results for 
all higher moments from the stochastic path integral formalism.

\section{Grand Canonical Treatment}

In order to see how the same result for the average condensate particle number
emerges from a modified grand canonical treatment, we recall that the grand
canonical formula for the average of the total particle number $N$ is \cite
{Huang}
\begin{equation}
N=\sum\limits_{k=0}^{\infty }\frac{1}{e^{\beta (\epsilon _{k}-\mu )}-1}%
=\sum\limits_{k=0}^{\infty }\langle n_{k}\rangle ,  \label{N gc
eq}
\end{equation}
where the sum runs over all states of energy $\epsilon _{k}$, $\beta
^{-1}=k_{B}T$ and the chemical potential $\mu $ is related to the mean
condensate particle number $\langle n_{0}\rangle $ as (we assume $\epsilon _{0}=0$)
\begin{equation}
\mu =-\beta ^{-1}\ln \left( \frac{1}{\langle n_{0}\rangle }+1\right) \text{.}
\label{mu <n0>}
\end{equation}
Various thermodynamic quantities of the gas such as the chemical potential $%
\mu $, the internal energy $U=\int \mu dN$ and the specific heat $C=dU/dT$
can be obtained analytically if we have an analytic expression for $\langle
n_{0}\rangle $. Equation (\ref{mu <n0>}) shows that $\mu $ is
nonzero for finite $N$, in contrast to the thermodynamic limit $%
N\rightarrow \infty $ where $\mu $ vanishes.

By using Eq. (\ref{mu <n0>}), $\langle n_{0}\rangle $ can be singled out
from Eq. (\ref{N gc eq}) as
\begin{equation}
N-\langle n_{0}\rangle =\sum\limits_{k>0}\frac{1}{\left( \frac{1}{\langle
n_{0}\rangle }+1\right) e^{\beta \epsilon _{k}}-1}\text{,}
\label{N gc singleout}
\end{equation}
which may be rewritten as
\begin{equation}
N-\langle n_{0}\rangle =\frac{1}{\left( \frac{1}{\langle n_{0}\rangle }%
+1\right) }\sum\limits_{k>0}\frac{1}{e^{\beta \epsilon _{k}}-\frac{\langle
n_{0}\rangle }{\langle n_{0}\rangle +1}}.  \label{N gc closed}
\end{equation}
This gives a self-consistent equation for the mean particle number $\langle
n_{0}\rangle $. In the limit $\langle n_{0}\rangle \gg 1$, the term $\langle
n_{0}\rangle /(\langle n_{0}\rangle +1)$ inside the summation may be
approximated by 1. As in the previous section, we denote the summation over $%
k$ as
\begin{equation}
\mathcal{H}=\sum_{k>0}\frac{1}{e^{\beta \epsilon _{k}}-1},  \label{calh}
\end{equation}
yielding a quadratic equation for the mean number of particles in the ground
state
\begin{equation*}
N-\langle n_{0}\rangle =\frac{\mathcal{H}}{\frac{1}{\langle n_{0}\rangle }+1}%
,
\end{equation*}
whose solution is
\begin{equation}
\langle n_{0}\rangle =\frac{1}{2}\left( N-\mathcal{H}-1+\sqrt{(N-\mathcal{H}%
-1)^{2}+4N}\right) ,  \label{n0 semianalytic2}
\end{equation}
which exactly coincides with (\ref{n0meapprox}), linking the two approaches.
It is natural to ask about the physical significance of this coincidence. The
master equation result is derived from a particle-number conserving formalism,
which already has the canonical ensemble property built in.  However, 
the master equation is solved approximately in Eq.~(\ref{n0meapprox}) 
under the condition of large condensate particle number. 
On the other hand, the result of the grand canonical analysis, Eq.~(\ref{n0 semianalytic2}),
is derived by trading the chemical potential for the average
particle number, which corresponds to imposing the total particle
number constraint on average. The resulting equation is further solved self-consistently, also
under the large condensate number assumption.  The coincidence of results
in this approximation, and the excellent agreement with the numerical 
canonical ensemble simulation, indicates that a strict accounting of
the canonical ensemble constraint is unnecessary for the calculation of $\la n_0\ra$.  

One should mention that the trick used in Eq. (\ref{N gc closed}) is
nontrivial. A straightforward expansion in $1/\langle n_{0}\rangle $ yields
very poor accuracy. We discuss this in Appendix A.

\section{Analytic expressions for mean condensate particle number and variance within
the generalized grand canonical approach}

\begin{figure}[t]
\center\epsfxsize=8.0cm\epsffile{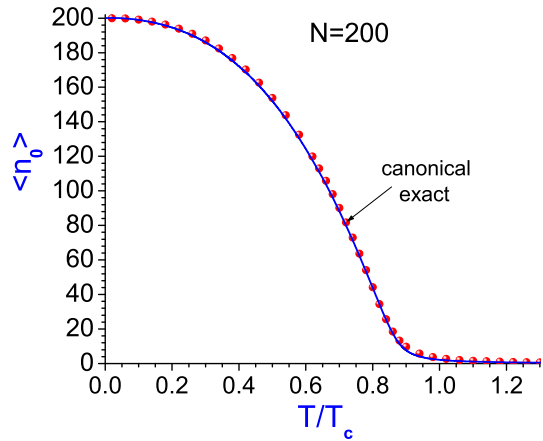}
\caption{(Color online) The average condensate particle
number is plotted versus temperature for $%
N=200$ particles in an isotropic harmonic trap.
The approximate analytic result given by Eqs. (\ref{n0meapprox}) (or
\ref{n0 semianalytic2}) and (\ref{analytic}) (solid line) is compared with
the ``exact canonical dots" computed numerically with good
agreement.}
\label{meanfig}
\end{figure}
\begin{figure}[b]
\center\epsfxsize=8.0cm\epsffile{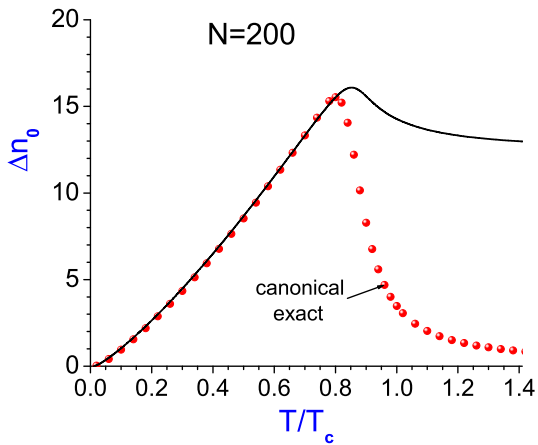}
\caption{(Color online) The variance of the condensate distribution is plotted versus
temperature for $N=200$. The approximate analytic result within the
generalized grand canonical approach (solid line) is compared with the
\textquotedblleft exact dots" computed numerically in the canonical
ensemble. The analytic result is in good agreement for low temperature, but
fails at temperatures comparable to the critical temperature and above. The
reason for the high-temperature discrepancy is that the modified grand
canonical theory neglects correlations between excited levels.}
\label{varfig}
\end{figure}
It is of interest to have an accurate analytic expression for the mean
number of particles in a mesoscopic condensate that is valid for all
temperatures. Let us specialize to the case of an isotropic harmonic trap.
Single-particle energy levels are then $\epsilon _{k}=\hbar \Omega (l+m+n)$,
where $\Omega $ is the trap frequency and $k=\{l,m,n\}$ denotes the quantum
numbers. In this case, the coefficient $\mathcal{H}$ can be evaluated
approximately by the method shown in the Appendix B. The idea of this method
is to first convert the triple sum into a single sum, and then to
approximate the single sum as an integral. We obtain
\begin{equation}
\mathcal{H}\approx \frac{\zeta \left( 3\right) }{a^{3}}+\left( \frac{\pi }{2a%
}\right)^{2}+\frac{1}{a}\left( \ln 2-\ln a\right) ,
\label{analytic}
\end{equation}
where $a=\beta \hbar \Omega $. The first term has been previously obtained
in Ref. \cite{GHcusp}, and here we find subdominant corrections in $a$.
Thus, we have an analytical expression for $\langle n_{0}\rangle $ given by
Eq. (\ref{n0 semianalytic2}), where the term $N-\mathcal{H}-1$ can be
expressed as
\begin{eqnarray}
&&N-\mathcal{H}-1\approx N\left\{ 1-\left( \frac{T}{T_{c}}\right)
^{3}\right\} \allowbreak -\frac{\pi ^{2}}{4}\left( \frac{N}{\zeta (3)}%
\right) ^{2/3}\left( \frac{T}{T_{c}}\right) ^{2}  \notag \\
&&-1+\left( \frac{N}{\zeta (3)}\right) ^{1/3}\frac{T}{T_{c}}\left\{ \ln %
\left[ \left( \frac{\zeta (3)}{N}\right) ^{1/3}\frac{T_{c}}{T}\right] -\ln
2\right\} ,  \label{M analytic}
\end{eqnarray}
and the critical temperature $T_{c}=(N/\zeta (3))^{1/3}\hbar \Omega /k_{B}$
has been introduced. For large $N$, the second line of (\ref{M analytic})
may be neglected, and if we also take $(N-\mathcal{H}-1)^{2}\gg 4N$, then
Eq. (\ref{n0 semianalytic2}) reduces to Eq. (8) of Ref.~\cite{GHcusp}:
\begin{equation}
\langle n_{0}\rangle \approx N\left\{ 1-\left( \frac{T}{T_{c}}\right)
^{3}\right\} \allowbreak -\left( \frac{N}{\zeta (3)}\right) ^{2/3}\left(
\frac{\pi }{2}\frac{T}{T_{c}}\right) ^{2},  \label{n0 GH}
\end{equation}
which further reduces to $\langle n_{0}\rangle \approx N\left[ 1-\left(
T/T_{c}\right) ^{3}\right] $ in the thermodynamic limit.

We now consider fluctuations of the number of particles in the condensate,
starting with the variance. The variance of the \textit{condensate}
particles can be expressed in terms of the \textit{non-condensate} variance
using the identity $n_{0}=N-\sum_{k>0}n_{k}$. In the grand canonical
approach, $\langle n_{k}^{s}\rangle $ can be evaluated from
\begin{equation}
\langle n_{k}^{s}\rangle =\frac{1}{Z_{k}}\sum\limits_{n_{k}}n_{k}^{s}e^{-%
\beta (E_{k}-\mu )n_{k}}=\frac{1}{Z_{k}}\beta ^{-s}\partial _{\mu }^{s}Z_{k},
\label{mean n^s}
\end{equation}
where $Z_{k}=(1-e^{-\beta (E_{k}-\mu )})^{-1}$. It follows from Eq. (\ref
{mean n^s}) that
\begin{equation}
\langle n_{k}^{2}\rangle =2\langle n_{k}\rangle ^{2}+\langle n_{k}\rangle .
\label{mean nk^2}
\end{equation}
We then obtain %\begin{widetext}
\begin{eqnarray}
\mu_2 &=& \langle (n_{0}-\langle n_{0}\rangle )^{2}\rangle \approx
\sum\limits_{k>0}[\langle n_{k}^{2}\rangle -\langle n_{k}\rangle ^{2}]
\notag \\
&\approx&\sum\limits_{k>0}[\langle n_{k}\rangle ^{2}+\langle n_{k}\rangle ]
\notag \\
&\approx&\sum_{k>0}\left\{ \frac{1}{(Ae^{\beta \epsilon _{k}}-1)^{2}} +
\frac{1}{Ae^{\beta \epsilon _{k}}-1}\right\},  \label{uuu}
\end{eqnarray}
where $A=1+1/\langle n_{0}\rangle$. We have assumed that fluctuations of $%
n_k $ ($k\ne 0$) are uncorrelated ($\langle n_k n_m \rangle=\langle n_k
\rangle \langle n_m \rangle$, $k\ne m$). This is the case provided $\langle
n_{0}\rangle $ is much larger then its variance, so that particle exchange
with the condensate reservoir is the main channel of fluctuations of $n_k$.
Near and above $T_c$, the correlations become substantial, which yields the
failure of Eq. (\ref{uuu}). Taking the chemical potential $\mu$ as given in (%
\ref{mu <n0>}) and using the method described in Appendix B, we find the
following analytic expression for the variance in the generalized grand
canonical analysis:
\begin{eqnarray}
\mu_2&\approx& \frac{7Ae^{a/2}+8}{8a(Ae^{a/2}-1)} + \frac{3}{2a^{2}}\ln
\left( \frac{A}{Ae^{a/2}-1}\right)  \label{Dn0analytic} \\
&+& \frac{1}{a^{3}}\left[ \frac{\pi ^{2}}{6}+\ln \left( \frac{Ae^{a/2}-1}{%
\sqrt{A}}\right) \ln A+\text{dilog}(Ae^{a/2})\right],  \notag
\end{eqnarray}
%\end{widetext}
where $a=\beta \hbar \Omega $ and di$\log (x)=\int_{1}^{x} dt \ln(t)/(1-t)$.
In the limit $k_{B}T\gg \hbar \Omega $ and $\langle n_{0}\rangle \gg 1$,
Eq.~(\ref{Dn0analytic}) yields $\mu_2 \approx \allowbreak
\pi ^{2}/6a^{3}=\pi ^{2}T^{3}N/6T_{c}^{3}\zeta (3)$ which agrees with Eq.
(11) of Ref.~\cite{pol}.

The analytic results are compared with numerical simulation in Figs. \ref
{meanfig}-\ref{varfig}. Fig. \ref{meanfig} shows an excellent agreement
between $\langle n_{0}\rangle $ computed analytically from Eqs. (\ref{n0
semianalytic2}) and (\ref{analytic}) and the exact numerical simulation
obtained in the canonical ensemble \cite{la}. In Fig. \ref{varfig} we plot
the variance $\Delta n_{0}$ as a function of temperature obtained from Eq. (%
\ref{Dn0analytic}) (solid line), as well as the \textquotedblleft exact
canonical dots" obtained from numerical computation \cite{la}. While the
analytic result is good at low temperature, at temperatures comparable to $%
T_{c}$ and above, there is substantial deviation from the canonical ensemble
result because correlations between excited levels are neglected.

\section{Analytic expression for all cumulants via the stochastic path
integral formalism}

While the master equation is a powerful approach, it is often the
case that a simple analytic solution cannot be found.
It is therefore of great interest to pursue
alternative treatments that give an approximate analytic solution of the
master equation, that is asymptotically valid in the physically relevant
limit of many particles in the condensate, $\langle n_{0}\rangle \gg 1$.
Just such an approach was developed in Refs.~\cite{anj1,anj2}, by solving
the fluctuation statistics problem with a stochastic path integral. The
stochastic path integral formalism is complimentary to the master equation
approach as will be shown below. However, we stress that it can also be
applied in cases where it is impossible to even write down a differential
master equation, as demonstrated in Ref.~\cite{anj2}.

The calculational details are given in Appendix C, but the basic idea is to
translate the master equation into the stochastic path integral, whose
action functional contains all rate information, and also imposes local
particle conservation. The fluctuation statistics can then be calculated in
saddle-point approximation by finding the ``zero energy lines" of the
dynamics - the statistical trajectory in phase space that is most likely,
similar to the instanton trajectory of Ref.~\cite{anjbi}. From this
trajectory, the generating function may be found as an area in phase space.

Rather than solving the original master equation (\ref{generalme}),
we skip to the generalization of CNB2 \cite{CNBII}, where
\begin{equation}
K_{n_{0}}=(N-n_{0})(1+\eta ),\qquad H_{n_{0}}=\mathcal{H}+(N-n_{0})\eta ,
\label{hk2}
\end{equation}
which applies also to higher temperatures, and $\eta $ is defined as
\begin{equation}
\eta =\mathcal{H}^{-1}\sum_{k>0}\frac{1}{(e^{\beta \epsilon _{k}}-1)^{2}}.
\label{eta}
\end{equation}
We introduce the notation $\kappa _{n}$ for the cumulants of the
distribution \cite{kum}, and define the nonstandard cumulant generating
function $Q(\lambda )$ as
\begin{equation}
\kappa _{s}=\partial _{\lambda }^{s-1}Q(\lambda )|_{\lambda =0}.
\label{simp}
\end{equation}
According to the calculations in Appendix C, this function is given by the
solution of the equation
\begin{equation}
(Q+1)K_{Q}(e^{\lambda }-1)+QH_{Q}(e^{-\lambda }-1)=0,  \label{zeroenergy}
\end{equation}
for {\it any} $K,H$. For the special case of (\ref{hk2}), the result is
\begin{figure*}[t]
\center\epsfxsize=16.0cm\epsffile{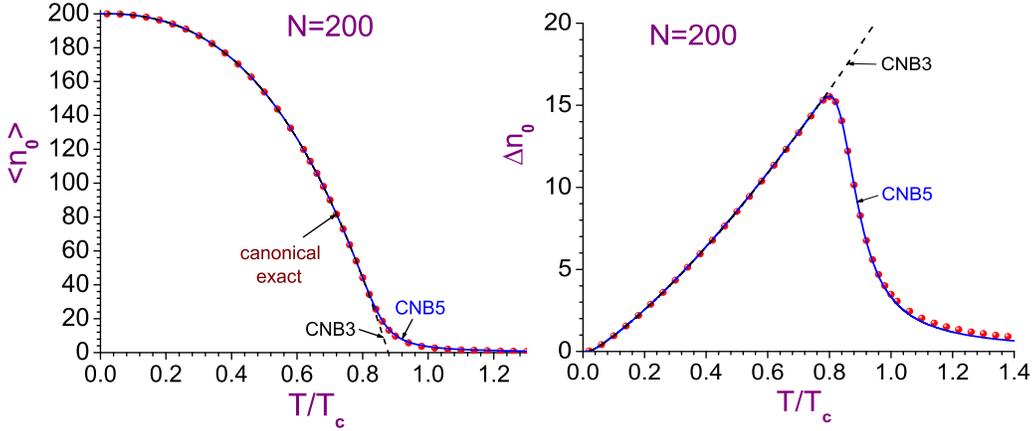} \caption{(Color online) The present
hybrid theory, we call CNB5, (solid line) is compared with the exact
numerical results obtained in the canonical ensemble
(dots). (left) The average condensate
particle number $\la n_0\ra$ and (right) the variance $\Delta n_0$
are plotted for $N=200$ versus temperature.  The predictions of
CNB3 \cite{CNBIII} (dashed line) are also shown.} \label{cond1}
\end{figure*}
\begin{widetext}
\be Q(\lambda) = \frac{-{\cal H} +(1+\eta)(N-1)e^\lambda - N \eta
+ \sqrt{4 e^\lambda (1+\eta) (-\eta + e^\lambda (1+\eta))N +
({\cal H} - e^\lambda (1+\eta)(N-1) +\eta N)^2}}{-2\eta + 2
e^{\lambda} (1+\eta)}. \label{Q} \ee
Applying Eq.~(\ref{simp}), the average value $(\kappa_1=\la n_0 \ra$) is given by
\be \langle n_0 \ra =(1/2) [(N- {\cal H} -1-\eta) + \sqrt{(N-{\cal
H} -\eta -1)^2 + 4 N (1+\eta)}], \label{ave} \ee
which coincides with (\ref{n0meapprox}) in the $\eta=0$ limit. The
second central moment $(\kappa_2 = \mu_2)$ is given by
\be
%\la n_0^2 \ra_c
\mu_2= \frac{(1+\eta)(\eta+{\cal H}) \sqrt{(N-{\cal H} -\eta -1)^2
+ 4 N (1+\eta)} - (1+\eta) [\eta^2 + {\cal H} (1+{\cal H} -N)
+\eta (1+2 {\cal H} + N)]}{2 \sqrt {(N-{\cal H} -\eta -1)^2 + 4 N
(1+\eta)}}, \label{second} \ee in terms of elementary functions.
\end{widetext}%\twocolumngrid
All higher cumulants and central moments may be easily computed from (\ref{Q}%
). Figures comparing these approximate results with the exact master
equation solution (shown in Fig. \ref{cond}) are not shown, simply
because for $N=200$, they are indistinguishable.

\section{Hybrid theory - combining CNB3 with the master equation analysis}

We now demonstrate how to combine ideas from the canonical ensemble 
quasiparticle formalism of CNB3 \cite{CNBIII} (which works well at low temperature when $\sqrt{\mu _{2}}\ll \bar{n%
}_{0}$) with the physics of the master equation approach, in order to obtain
essentially perfect quantitative agreement with the exact numerical solution
of the canonical partition function at \textit{all} temperatures for the
fluctuation statistics of the Bose gas.

Defining the function $F_{n_{0}}$ as the ratio between the probabilities to
find $n_{0}+1$ and $n_{0}$ particles in the ground state,
\begin{equation}
F_{n_{0}}=\frac{p_{n_{0}+1}}{p_{n_{0}}},  \label{F0}
\end{equation}
we note that the canonical ensemble constraint is imposed by $F_{N}=0$
because if all the particles are in the condensate, it is impossible to cool
further, or, in other words, the probability to find $N+1$ particles in the
condensate is equal to zero. It is then useful to consider an expansion of
this function in $N-n_{0}$. Rather than Taylor expand, a better approach is
to approximate this function by ratio of two power series and then determine
both the numerator and denominator coefficients, a procedure known as a
Pad\'{e} approximation. Pad\'{e} approximations are usually superior to
Taylor expansions when functions contain poles, because the use of rational
functions allows them to be well-represented. We approximate
\begin{equation}
F_{n_{0}}=\frac{K_{n_{0}}}{H_{n_{0}+1}},  \label{F}
\end{equation}
where the functions $H,K$ are both polynomials in $N-n_{0}$,
\begin{eqnarray}
H_{n_{0}} &=&\mathcal{H}+\eta (N-n_{0})+\alpha (N-n_{0})^{2},  \notag \\
K_{n_{0}} &=&(1+\eta )(N-n_{0})+\alpha (N-n_{0})^{2},  \label{heat}
\end{eqnarray}
and truncate the expansion at second order. Knowledge of the function $%
F_{n_{0}}$ allows the construction of the entire distribution,
\begin{equation}
p_{n_{0}}=Z_{N}^{-1}\prod_{m=n_{0}}^{N-1}F_{m}^{-1},\qquad
Z_{N}=\sum_{n_{0}=0}^{N}\prod_{m=n_{0}}^{N-1}F_{m}^{-1},  \label{dist}
\end{equation}
or
\begin{equation}
p_{n_{0}}=C\frac{(N-n_{0}-1+x_{1})!(N-n_{0}-1+x_{2})!}{%
(N-n_{0})!(N-n_{0}+(1+\eta )/\alpha )!},  \label{p4}
\end{equation}
where $x_{1,2}=(\eta \pm \sqrt{\eta ^{2}-4\alpha \mathcal{H}})/2\alpha $ and
$C$ is the normalization constant determined by $\sum\limits_{n_{0}=0}^{N}$ $%
p_{n_{0}}=1$. The functions $H,K$ take the same form as in the master
equation, but now the coefficients $\mathcal{H},\eta ,\alpha $ are treated
as free parameters to be fixed by comparison with CNB3 \cite{CNBIII} at low
temperatures. %In order to justify the expansion
%(\ref{F},\ref{heat}), we numerically plot $F_{n_0}$ as a function
%of $n_0$ for several values of temperature.  In all cases, the low
%$n_0$ plateau of this function is better captured by a Pad\'{e}
%approximation, over a Taylor expansion.
%

The further analytic input for the theory is the first three moments of the
distribution described by the heating and cooling coefficients (\ref{heat})
in the low temperature limit. These moments are used to fix the free
parameters $\mathcal{H},\eta ,\alpha $. The calculation is done in Appendix
C for the complete generating function using the stochastic path integral
formalism, but here we only reproduce the needed first three:
\begin{eqnarray}
\langle n_{0}\rangle &=&N-\mathcal{H},  \label{fn0} \\
\mu _{2} &=&\mathcal{H}(1+\eta +\alpha \mathcal{H}), \\
\mu _{3} &=&-\mathcal{H}(1+\eta +\alpha \mathcal{H})(1+2\eta +4\alpha
\mathcal{H}).  \label{first3}
\end{eqnarray}
Comparison with the CNB3 \cite{CNBIII} at low temperature allows us to
obtain the parameters $\mathcal{H},\eta ,\alpha $,
\begin{figure*}[t]
\center\epsfxsize=18.0cm\epsffile{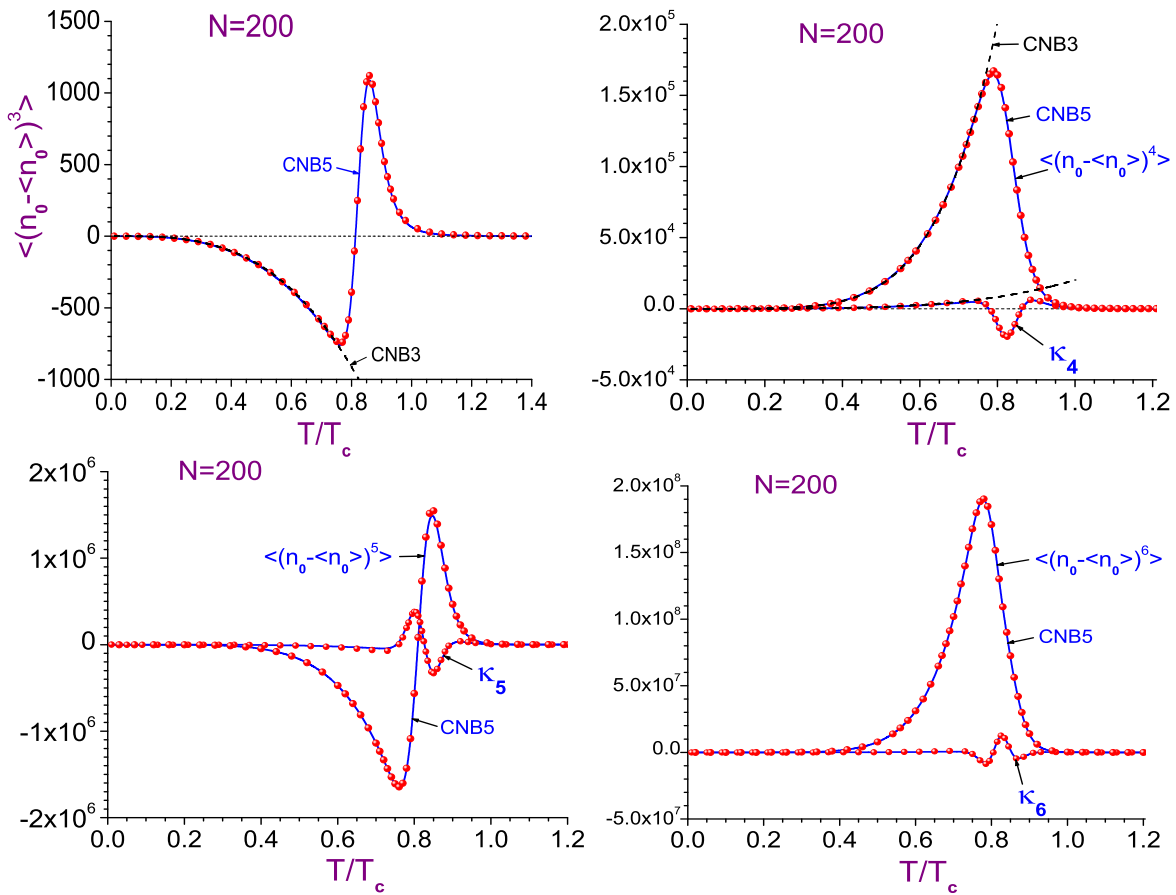} \caption{(Color online) Same
as in Fig. \ref{cond1}, for the third, fourth, fifth and sixth centered
moments, denoted by $\la (n_0 -\la n_0\ra)^n\ra$, as well as
cumulants $\kappa_n$ \cite{kum}. } \label{cond2}
\end{figure*}
\begin{widetext}
\onecolumngrid
\bea {\cal H} &=& \sum_{k\ne 0} {\bar n_k}, \qquad \eta =
\frac{1}{2} \left( -\frac{\sum_{k\ne 0} (2 {\bar n}_k^3 + 3 {\bar
n}_k^2 + {\bar n}_k)}{\sum_{k\ne 0} ({\bar n}_k^2 + {\bar n}_k)}
-3 + \frac{4 \sum_{k\ne 0}  ({\bar n}_k^2 + {\bar
n}_k)}{\sum_{k\ne 0} {\bar n}_k}\right), \nonumber \\
\alpha &=&  \frac{1}{\sum_{k \ne 0} {\bar n}_k } \left(
\frac{1}{2} -\frac{ \sum_{k\ne 0}  ({\bar n}_k^2 + {\bar
n}_k)}{\sum_{k\ne 0} {\bar n}_k} +\frac{\sum_{k\ne 0} (2 {\bar
n}_k^3 + 3 {\bar n}_k^2 + {\bar n}_k)}{2\sum_{k\ne 0} ({\bar
n}_k^2 + {\bar n}_k)} \right),
\label{uuu1}
\eea
\end{widetext}
where ${\bar{n}}_{k}=(e^{\beta \epsilon _{k}}-1)^{-1}$. Knowing these
parameters allows the complete specification of the entire distribution in
this approximation. To sum up: The theory uses ($i$) the master equation/stochastic path integral 
formalism to determine the first three moments at low temperature, and ($ii$)
the results of CNB3 (that works well in the low temperature limit) to fix
the three undefined parameters. The distribution function (\ref{p4})
together with Eqs. (\ref{uuu1}) then gives predictions for \textit{all}
central moments and cumulants \cite{kum} at \textit{all} temperatures.

The theory is put to the test in Figs. \ref{cond1} and \ref{cond2} for the
first few central moments and cumulants of a thermal Bose gas in a harmonic
trap, for the mesoscopic case $N=200$. The hybrid treatment yields perfect
agreement with the ``exact" numerical dots obtained in the canonical
ensemble.

\section{Conclusions}

We have discussed the fluctuation statistics of an ideal, mesoscopic, Bose-Einstein condensate
from several different perspectives. First, we have reviewed the master
equation approach, and derive an approximate analytic solution for the mean
condensate particle number. By generalizing the grand canonical analysis, the same
result from the approximate solution of a self-consistent equation has been
recovered for the mean condensate particle number. Improved analytic results are
obtained for the mean condensate particle number in the case of a 3D harmonic trap,
that are quite accurate when compared with numerical calculation of the
canonical partition function result. Analogous treatment of the variance of
the distribution in the generalized grand canonical picture have given
results that are very accurate below the critical temperature, but
substantially deviate at or above the critical temperature. The reason for
this discrepancy is because correlations of excited energy levels were
neglected in the calculation.

Next, we have presented an (approximate) analytic solution for the
generating function of the fluctuation statistics from the master equation
perspective. This is done by employing the stochastic path integral
formalism, with the saddle-point approximation, giving results that are
asymptotically valid in the physically relevant limit of many particles in
the condensate. The general solution is discussed for arbitrary heating and
cooling coefficients, and specific results are given in terms of elementary
functions for the Bose-Einstein condensate heating and cooling coefficients
of CNB2. These results are in excellent agreement with
exact master equation solution when compared numerically.

A hybrid theory has been put forth that combines the master
equation/stochastic path integral with the approach of CNB3. The theory
applies the results of CNB3 in the low temperature limit, together with the
predictions of the master equation/stochastic path integral. This is
accomplished by preserving the physical structure of the master equation,
while using the first three moments of CNB3 to fix the numerical values of
the heating and cooling parameters. These predictions are then examined for
several higher moments (or cumulants), at all temperatures. The predictions
of this theory are essentially in perfect agreement with numerical
simulation of the exact canonical partition function.

Finally, we briefly discuss how our methods and results extend in the presence of weak interactions.  As will be shown in Ref.~\cite{anatoly}, it is easy to generalize the hybrid approach to the interacting case and take the first three central moments (in the low-temperature limit) from CNB3.  From the microscopic master equation point of view, interactions generally lead to ``off-diagonal'' transitions in the density matrix, demanding a fully coherent treatment of the problem.  However, it will be demonstrated elsewhere that good agreement with CNB3 (in the applicable low temperature limit) may be obtained with only diagonal transitions, where the heating and cooling coefficients are determined from Bogolioubov theory.  

\section{Acknowledgments}

We wish to thank M.O. Scully for the initiation of our work and stimulating
discussions. We gratefully acknowledge the support of the Office of Naval
Research (Award No. N00014-03-1-0385) and the Robert A. Welch Foundation
(Grant No. A-1261).

\appendix

\section{Grand canonical formalism: simple expansion with triple summation}

\begin{figure}[t]
\center\epsfxsize=8.0cm\epsffile{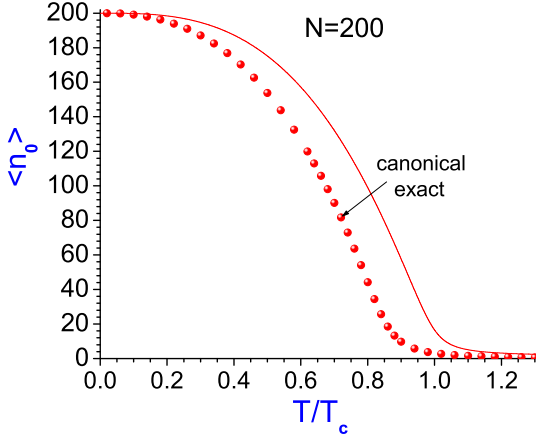}
\caption{(Color online) The average condensate particle number is plotted versus
temperature for $N=200$. The approximate analytic result (\ref{n0 Scully})
within the grand canonical approach (solid line) is compared with the
\textquotedblleft exact canonical dots". The triple sum, together with the
simple expansion gives a poor fit.}
\label{meanfigbad}
\end{figure}

Before calculating the analytic expressions for the average and variance of
the condensate fluctuations, it is instructive to first take another path,
different than the one presented in Sec. III. Rather than make the step (\ref
{N gc closed}), we make the simple expansion
\begin{eqnarray}
N-\langle n_{0}\rangle &=& \sum\limits_{r=1}^{\infty }\sum\limits_{k>0}\frac{%
1}{(e^{\beta \epsilon _{k}}-1)^{r}}\left( -\frac{e^{\beta \epsilon _{k}}}{%
\langle n_{0}\rangle }\right) ^{r-1}  \notag \\
&\simeq& A(T)-\frac{1}{\langle n_{0}\rangle }B(T),
\label{n0
expand wrong 2nd}
\end{eqnarray}
where
\begin{equation}
A(T) =\sum\limits_{k>0}\frac{1}{e^{\beta \epsilon _{k}}-1}, \quad B(T)
=\sum\limits_{k>0}\frac{e^{\beta \epsilon _{k}}}{(e^{\beta \epsilon
_{k}}-1)^{2}}.
\end{equation}
The above expansion is in the small parameter $e^{\beta \epsilon _{k}}/
[(e^{\beta \epsilon_k}-1) \langle n_{0}\rangle]$. Equation (\ref{n0 expand
wrong 2nd}) has the solution
\begin{equation}
\langle n_{0}\rangle =\frac{1}{2}\left( N-A+\sqrt{(N-A)^{2}+4B}\right) .
\end{equation}
For an isotropic harmonic trap, $\epsilon _{k}=\hbar \Omega (l+m+n)$, and
since $e^{-a(l+m+n)}<1$ (here $a=\beta \hbar \Omega $) we make the series
expansion
\begin{eqnarray}
A(T) &=& \sum\limits_{l,m,n=1}^{\infty }\frac{e^{-a(l+m+n)}}{1-e^{-a(l+m+n)}}
\notag \\
&=& \sum\limits_{l,m,n=1}^{\infty }\sum\limits_{s=0}^{\infty
}e^{-(s+1)a(l+m+n)}  \notag \\
&=&\sum\limits_{s=0}^{\infty }\left[ \sum\limits_{n=1}^{\infty }e^{-(s+1)an}%
\right] ^{3}=\sum\limits_{s=1}^{\infty }\left[ \sum\limits_{n=1}^{\infty
}e^{-san}\right] ^{3}  \notag \\
&\simeq& \sum\limits_{s=1}^{\infty }\left[ \frac{1}{a}\int_{0}^{\infty
}e^{-sz}dz\right] ^{3}\rightarrow \frac{1}{a^{3}}\zeta (3)\text{,}
\label{Atriple}
\end{eqnarray}
where $\zeta (n)=\sum\limits_{s=1}\frac{1}{s^{n}}$ is the Riemann-zeta
function and the conversion to integration is based on the assumption $a< 1 $%
. Similarly we find %\begin{widetext}
\begin{eqnarray}
B(T) &=&\sum\limits_{l,m,n=1}^{\infty }\frac{e^{a(l+m+n)}}{%
[e^{a(l+m+n)}-1]^{2}}  \notag \\
&=&\sum\limits_{l,m,n=1}^{\infty }\sum\limits_{s=0}^{\infty
}(s+1)e^{-(s+1)a(l+m+n)}  \notag \\
&=&\sum\limits_{s=0}^{\infty }(s+1)\left[ \sum\limits_{n=1}^{\infty }
e^{-(s+1)an}\right] ^{3}  \notag \\
&=&\sum\limits_{s=1}^{\infty }s\left[ \frac{1}{a}\int_{a}^{\infty }e^{-sz}dz%
\right] ^{3} =\frac{1}{a^{3}}\zeta (2).  \label{Btriple}
\end{eqnarray}
Finally, by noting that $A(T)=N(T/T_{c})^{3},$ $B=N(T/T_{c})^{3}\zeta
(2)/\zeta (3)$, where $T_{c}=[N/\zeta (3)]^{1/3}\hbar \Omega /k_{B}$ is
critical temperature in the thermodynamic limit, we obtain
\begin{eqnarray}
\langle n_{0}\rangle &=&\frac{N}{2}\left\{ 1-\left( \frac{T}{T_{c}}\right)
^{3}\right\}  \notag \\
&+&\frac{N}{2}\sqrt{\left[ 1-\left( \frac{T}{T_{c}}\right) ^{3}\right] ^{2}+%
\frac{\zeta (2)}{\zeta (3)}\frac{4}{N}\left( \frac{T}{T_{c}}\right) ^{3}}.
\label{n0 Scully}
\end{eqnarray}
%\end{widetext}
For large $N$, Eq. (\ref{n0 Scully}) reproduces the thermodynamic limit.
However, the triple sum formula is inaccurate for small $N$ especially near $%
T_{c}$. Result (\ref{n0 Scully}) is plotted in Fig.~\ref{meanfigbad}. The
fit is poor because $N=200$ is not sufficiently large. We note that this
expansion may be systematically improved by keeping more terms in the $%
1/\langle n_0\rangle$ expansion, and evaluating the coefficients in similar
manner. In the next appendix, this analysis is improved.

\section{Grand canonical formalism: improved expansion with single summation}

The goal of this appendix is to obtain an improved analytic expression for
both the average and variance of the condensate distribution for an
isotropic harmonic trap. Equation (\ref{n0 expand wrong 2nd}) is obtained
from Eq. (\ref{N gc singleout}) by neglecting the terms with $1/\langle
n_{0}\rangle^{2}$ and higher under the condition $\langle n_{0}\rangle \gg
e^{\beta \epsilon _{k}}/(e^{\beta \epsilon _{k}}-1)$. For the enhanced
expansion (\ref{N gc closed}), it is not difficult to see that the validity
condition is $\langle n_{0}\rangle +1 \gg (e^{\beta \epsilon _{k}}-1)^{-1}$,
allowing for a better approximation.

After this point, there are two more key steps: (1) convert the triple
summation into a single summation, (2) make an integral approximation to the
single summation. We first recall
\begin{equation}
\mathcal{H}=\sum_{k>0}\frac{1}{(e^{\beta \epsilon _{k}}-1)},  \label{H}
\end{equation}
where for a harmonic trap $\beta \epsilon _{k}=a(l+m+n)$ and we have
introduced $a=\hbar \Omega \beta $. The triple summation can be reduced to a
single summation over $s$, where $s=l+m+n$, and weighting this sum with the
number of ways $W$ to put $s$ quanta into three boxes, $%
W=(s+2)!/(s!2!)=(s+2)(s+1)/2$,
\begin{equation}
\mathcal{H}=\sum_{s=1}^{\infty }\frac{(s+2)(s+1)}{2(e^{as}-1)}.  \label{H1}
\end{equation}

In order to find an analytical expression for $\mathcal{H}$, we interpret
the summation as a Riemann summation, and convert it (approximately) to an
integral using the mid-point rule, $\sum_{s=1}^{\infty }f_{s} \approx
\int_{0}^{\infty }ds f(s+1/2)$. The midpoint rule gives the better
approximation because it compromises between the lower and upper summation.
Re-parameterizing the integral yields
\begin{eqnarray}
\mathcal{H} &\approx &\frac{1}{2}\int_{a/2}^{\infty }\left( \frac{x^{2}}{%
a^{2}}+\frac{3x}{a}+2\right) \frac{1}{e^{x}-1}\frac{dx}{a}  \notag \\
&=&\frac{\zeta \left( 3\right) }{a^{3}}-\frac{2\ln \frac{4}{a+4}}{a^{3}}-%
\frac{1}{2a^{2}}  \notag \\
&&+\allowbreak \frac{\pi ^{2}}{4a^{2}}+\frac{3}{2a^{2}}\text{dilog}\left( e^{%
\frac{1}{2}a}\right) +\frac{3}{16}  \notag \\
&&+\frac{1}{2}-\frac{1}{a}\ln \left( e^{a/2}-1\right)  \label{Aanalytic} \\
&\approx &\frac{\zeta \left( 3\right) }{a^{3}}+\left( \frac{\pi }{2a}\right)
\allowbreak ^{2}+\frac{1}{a}\left( \ln 2-\ln a\right),
\end{eqnarray}
where di$\log (x)=\int_{1}^{x} dt \ln(t)/(1-t)$.
%where we have used $\int_{a}^{\infty }\frac{dx}{e^{x}-1}=a-\ln
%\left( e^{a}-1\right) $, $\int_{0}^{\infty }\frac{xdx}{e^{x}-1}=\allowbreak \frac{%
%\pi ^{2}}{6}$ and $\int_{0}^{\infty
%}\frac{x^{2}dx}{e^{x}-1}=\allowbreak 2\zeta \left( 3\right) $.
This derivation gives Eq.~(\ref{analytic}), one of our main results.

Turning to the variance, we proceed in the same manner to obtain an
analytical expression for $\mu_2$. The conversion from the triple sum to the
single sum yields

%\begin{widetext}
\begin{equation}
\mu _{2}=\sum\limits_{s=1}^{\infty }\frac{\left( s+2\right) \left(
s+1\right) }{2}\left\{ \frac{1}{(Ae^{as}-1)^{2}}+\frac{1}{Ae^{as}-1}\right\},
\label{Dn0
1D}
\end{equation}
where $A=1+1/\langle n_0\rangle$. Next, we convert the sum into an
integral as before to yield
\begin{equation}
\mu _{2} \approx \int_{a/2}^{\infty }\frac{Ae^{x}}{(Ae^{x}-1)^{2}}\frac{1}{2}%
\left( \frac{x^{2}}{a^{2}}+3\frac{x}{a}+2\right) \frac{dx\allowbreak
\allowbreak \allowbreak }{a}.  \label{Dn0 integ}
\end{equation}
%
%where the value of $\langle n_{0}\rangle $ as a function of
%temperature can be computed from Eqs. (\ref{n0 semianalytic}) and
%(\ref{Aanalytic}) in the previous section.
Then we integrate by parts to find
\begin{equation}
\mu _{2}\approx\frac{15}{8a(Ae^{a/2}-1)}+\frac{1}{2a^{2}}\int_{a/2}^{\infty }\frac{%
\frac{2x}{a}+3}{Ae^{x}-1}dx.  \label{a4}
\end{equation}
The integral in Eq. (\ref{a4}) can be calculated analytically to find
\begin{eqnarray}
\mu _{2} &\approx&\frac{7Ae^{a/2}+8}{8a(Ae^{a/2}-1)}\allowbreak +\frac{3}{%
2a^{2}}\ln \left( \frac{A}{Ae^{a/2}-1}\right)  \label{Dn0analytic2} \\
&+&\frac{1}{a^{3}}\left[ \frac{\pi ^{2}}{6}+\ln \left( \frac{Ae^{a/2}-1}{%
\sqrt{A}}\right) \ln A+\text{dilog}(Ae^{a/2})\right] .  \notag
\end{eqnarray}
This derivation gives Eq.~(\ref{Dn0analytic}) another main result.

\section{Stochastic path integral solution of the master equation}

The purpose of this appendix is to provide approximate expressions for all
cumulants of the stationary condensate fluctuations using elementary
functions, starting from the master equation approach. In order to
accomplish this, we employ the stochastic path integral formalism \cite{anj1,anj2}.

Consider a general differential master equation of the form
\begin{equation}
{\dot{P}}_{n}(t)=\sum_{m}[W_{nm}P_{m}(t)-W_{mn}P_{n}(t)],  \label{cme}
\end{equation}
where $W_{nm}$ is a transition rate from $m$ to $n$, and $P_{n}$ is the
probability of occupying state $n$. Applying this equation to our BEC
problem, the microscopic rates to different states are taken from CNB2 \cite
{CNBII}:
\begin{eqnarray}
W_{n_{0},n_{0}-1} &=&\kappa (1+\eta )(N+1-n_{0})n_{0},  \notag \\
W_{n_{0},n_{0}+1} &=&\kappa \lbrack \mathcal{H}+\eta (N-n_{0}-1)](n_{0}+1),
\label{mrates}
\end{eqnarray}
where $n_{0}$ is the number of particles in the condensate, $\kappa $ is a
rate constant, $\mathcal{H}$ is given in (\ref{calh}) and $\eta $ is given
in (\ref{eta}). We now express this master equation as a stochastic path
integral, by going to a continuous representation where the discrete number
of particles in the ground state $n_{0}$ is replaced by an effectively
continuous variable $Q$, %\label{pathint}
\begin{eqnarray}
U(Q_{f},Q_{i},t) &=&\int \!\mathcal{D}Q\mathcal{D}\lambda \exp \{S(Q,\lambda
)\},  \label{path1} \\
S(Q,\lambda) &=&\int_{0}^{t}dt^{\prime }[-\lambda \dot{Q}+H(Q,\lambda )].
\label{action1}
\end{eqnarray}
The object $U$ is the evolution operator going from one particle
configuration $Q_{i}$ to another particle configuration $Q_{f}$ in time $t$.
It is expressed as a path integral over $Q$ and $\lambda $. The auxiliary
variable $\lambda $ is a canonically conjugate variable and imposes local
particle number conservation. In the continuous limit, (and suppressing the
overall rate constant $\kappa$) the CNB2 \cite{CNBII} rates may be expressed
as
\begin{eqnarray}
W(Q^{\prime },Q) &=&(1+\eta )(N+1-Q^{\prime })Q^{\prime }\delta (Q^{\prime
}-Q-1)  \label{conrates} \\
&+&[\mathcal{H}+\eta (N-Q^{\prime }-1)](Q^{\prime }+1)\delta (Q^{\prime
}-Q+1).  \notag
\end{eqnarray}
According to the prescription of Ref.~\cite{anj2}, the Hamiltonian $%
H(Q,\lambda )$ of the stochastic path integral is found from the equation:
\begin{equation}
H(Q,\lambda )=\int dQ^{\prime }\left[ e^{(Q^{\prime }-Q)\lambda }-1\right] W(%
{Q^{\prime }},{Q}).  \label{translation}
\end{equation}
For the rates (\ref{conrates}), we find
\begin{eqnarray}
H(Q,\lambda ) &=&(1+\eta )(N-Q)(Q+1)(e^{\lambda }-1)  \notag \\
&+&[\mathcal{H}+\eta (N-Q)]Q(e^{-\lambda }-1),  \label{hamagain}
\end{eqnarray}
or for the general master equation (\ref{generalme}) with arbitrary
coefficients $H,K$, we find
\begin{equation}
H(Q,\lambda )=K_{Q}(Q+1)(e^{\lambda }-1)+H_{Q}\,Q(e^{-\lambda }-1).
\label{hamagain1}
\end{equation}
This result has a simple physical interpretation: On a short time scale, the
elementary transitions into and out of the condensate are Poissonian,
witnessed by the generators $\mathcal{G}$ of Poissonian statistics, $%
\mathcal{G} = \Gamma [\exp(\pm \lambda) -1]$ (counting an incoming $(+)$ or
outgoing $(-)$ boson). These boson transitions are described with a rate into
the condensate $\Gamma_{\mathrm{in}} = K_{Q}(Q+1)$, and a rate out of the
condensate $\Gamma_{\mathrm{out}} = H_{Q}\,Q$.

The stochastic path integral (\ref{path1}) may be evaluated in saddle point
approximation, where the large parameter of the expansion is $\langle
n_{0}\rangle \gg 1$, the number of particles in the condensate. Applying
this approximation gives the analog of Hamilton's equations of motion,
\begin{equation}
{\dot{Q}}=\partial _{\lambda }H,\qquad {\dot{\lambda}}=-\partial _{Q}H.
\label{eom}
\end{equation}

To solve the problem of instantaneous particle number statistics in this
approximation, we generalize the method of Ref.~\cite{anj2}, following the
method of Ref.~\cite{anjbi}, by first finding the ``zero energy lines",
implicitly defined by the equation $H(Q,\lambda )=0$. For time scales longer
than the relaxation time $\kappa^{-1}$, any $Q$-distributed initial state
will be projected onto the zero energy lines. The trivial zero energy line
is given by $\lambda _{0}=0$ and must exist for the probability distribution
to be normalized. The instantaneous fluctuation statistics (to leading
order) can be found by calculating the statistical action (\ref{action1})
along the non-trivial zero energy line. This action, $S(\chi )$, is also the
generating function of the cumulants of the fluctuation statistics. On the
zero energy line, the Hamiltonian vanishes, leaving only the dynamical part
of the action,
\begin{equation}
S(\chi )=-\int_{0}^{t}dt^{\prime }{\lambda }(t^{\prime }){\dot{Q}}(t^{\prime
})=\int_{0}^{\chi }Q(\lambda )\,d\lambda ,  \label{act}
\end{equation}
and we have changed variables from time to phase space coordinates. In the
case of Bose-Einstein condensation, (\ref{hamagain}), the nontrivial
zero energy line $Q(\lambda )$ is given by
\begin{widetext}
\be Q(\lambda) = \frac{-{\cal H} +(1+\eta)(N-1)e^\lambda - N \eta
+ \sqrt{4 e^\lambda (1+\eta) (-\eta + e^\lambda (1+\eta))N +
({\cal H} - e^\lambda (1+\eta)(N-1) +\eta N)^2}}{-2\eta + 2
e^{\lambda} (1+\eta)}, \label{Q2} \ee
\end{widetext} In order to have a generating function, it is unnecessary to
perform the integral (\ref{act}) because $dS/d\chi =Q(\chi )$. Therefore,
all cumulants of the distribution can be found from %
%\be \la n_0^s \ra_c
\begin{equation}
\kappa _{s}=\partial _{\lambda }^{s-1}Q(\lambda )|_{\lambda =0}.
\label{simp2}
\end{equation}
This result generalizes the discussion of Ref.~\cite{anj2} for any two
Poissonian processes in series (equilibrium or not), and is easily
generalized to arbitrary elementary processes. Equations (\ref{Q2}) and (\ref
{simp2}) recover Eqs.~(\ref{simp}) and (\ref{Q}) and are main results.

In section VI, this same method is used to calculate the cumulants where the
heating and cooling coefficients are given by
\begin{eqnarray}
K_{n_{0}} &=&(1+\eta )(N-n_{0})+\alpha (N-n_{0})^{2},  \notag \\
H_{n_{0}} &=&\mathcal{H}+\eta (N-n_{0})+\alpha (N-n_{0})^{2},  \label{expand}
\end{eqnarray}
As before, we define $Q(\lambda )$ as the solution of the approximate zero
energy equation, $e^{\lambda }=H_{Q}/K_{Q}$, where the low temperature limit
allows the approximation $Q+1\rightarrow Q$. The solution is
\begin{widetext}
\be Q(\lambda) = N - \frac{\eta - e^\lambda(1+\eta) + \sqrt{4
\alpha {\cal H} (e^\lambda -1) + [\eta - (1+ \eta)e^\lambda]^2}}{2
\alpha (e^\lambda -1)}. \label{qgen} \ee
\end{widetext}%\onecolumngrid
The first three cumulants are given in Eqs.~(\ref{fn0})-(\ref{first3}),
which coincide with the first three central moments.

We also briefly note that time-averaged fluctuation statistics may be easily
calculated within the stochastic path integral formalism. Consider a
detector that has finite time resolution. The physical quantity that is of
interest is then the condensate particle number, averaged over some time window $\tau$%
,
\begin{equation}
Q_\tau = (1/\tau) \int_0^\tau dt^{\prime}n_0(t^{\prime}),  \label{aveQ}
\end{equation}
where we take $\tau$ longer than any dynamical time scale, for simplicity.
Following the method of Ref.~\cite{anj2}, we find the distribution $%
P(Q_\tau) $ is approximately given by the expression
\begin{equation}
\log P(Q_\tau) = - \tau [\sqrt{\Gamma_{\mathrm{in}}} - \sqrt{\Gamma_{\mathrm{%
out}}}]^2,  \label{anstimeave}
\end{equation}
where $\Gamma_{\mathrm{in}} = K_{Q_\tau}(Q_\tau+1)$, $\Gamma_{\mathrm{out}}
= H_{Q_\tau}\,Q_\tau$. Interestingly, this result is of the same form as the
time-averaged electron fluctuations on a mesoscopic cavity, out of
equilibrium \cite{anj2}. This similarity of statistics for radically
different physical systems originates from the fact that both systems can be
described as two Poissonian processes in series.


\begin{thebibliography}{99}
\bibitem{kum}  The cumulants $\kappa _{m}$ are defined as coefficients in
Taylor expansion $\ln \Theta _{n}(u)=\sum_{m=1}^{\infty }\kappa
_{m}(iu)^{m}/m!$, where $\Theta _{n}(u)$ is the characteristic function $%
\Theta _{n}(u)=\mathrm{Tr}\left\{ e^{iu\hat{n}}\hat{\rho}\right\} $. There
are simple relations between $\kappa _{m}$ and the central moments $\mu _{m}$%
, in particular, $\kappa _{1}=\bar{n}$, $\kappa _{2}=\mu _{2}$, $\kappa
_{3}=\mu _{3}$, $\kappa _{4}=\mu _{4}-3\mu _{2}^{2}$, $\kappa _{5}=\mu
_{5}-10\mu _{2}\mu _{3}$ and $\kappa _{6}=\mu _{6}-15\mu _{2}(\mu _{4}-2\mu
_{2}^{2})$. For a Gaussian distribution $\kappa _{m}=0$, for $m=3,4,\ldots $.
\end{thebibliography}
\end{document}